 \documentclass[preprint2]{aastex}
 \usepackage{amsmath,alltt}
\newcommand{\beq}[1]{\begin{eqnarray}\label{#1}}
\newcommand{\eeq}{\end{eqnarray}}

\slugcomment{}

 \shorttitle{SCMI}
 \shortauthors{Zeng \& Gao}

\begin{document}


 \title{Spherical Collapse Model And Dark Energy(II)}

 \author{Ding-fang Zeng \& Yi-hong Gao}
 \affil{Institute of Theoretical Physics, Chinese Academy of Science}
 \email{dfzeng@itp.ac.cn \& gaoyh@itp.ac.cn}

 \begin{abstract}
 This is a second paper of a series of two. In this paper, we
 directly correct the problem pointed out
 in the first paper of this series, dark energy does not cluster
 on the scale of galaxy clusters, but the current describing the
 flowing of dark energies outside the clusters is ignored in almost
 all the previous papers. We set up and solve
 a first order differential equation which
 describes the evolution of the clusters in a back ground
 universe containing dark energies.
 From the solution we extract the key parameters of the model
 and find them depending on the equation of state coefficients
 of dark energies rather non-trivially. We then apply the results
 in Press-Scheter theory and calculate the number density of
 galaxy clusters and its evolutions, we find the observable quantities are
 strongly affected by the equation of state coefficients of dark
 energies.
 \end{abstract}



 \maketitle

 \section{Introduction}

 In a previous paper \citep[]{SphereI},
 we point out that, in most of the existing
 literatures about top-hat spherical collapse model of galaxy clusters
 formation, \citep[]{Barrow, ECF, WangSteinhardt1, LokasHoffman, Peacock, NWeinberg},
 the dark energy is assumed not to cluster on the scale of
 galaxy clusters. But the current which describes the flowing of
 dark energy outside the clusters is usually ignored, so the discussions
 in these literatures are not self-consistent.
 In that paper, by assuming that dark energy clusters synchronously
 with ordinary matters so that the dark energy current flowing
 outside the clusters does not exist at all, we make our
 discussions self-consistent. However,
 by so doing, we do not correct
 the problem in the existing literatures directly. We only
 from the contrary indicate that the effects of the
 dark energy current may be important.

 The purpose of this paper is to directly consider the effects of
 such a current. Just as we state in that paper, when we add such
 a current in the energy-momentum tensor, Einstein equation
 becomes complicated and the metric of the cluster-inside
 space-time
 cannot be factorized as usual. So not only Friedman equation, but
 also Raycharduri equation does not follow as thought by the author
 of \citep[]{WangSteinhardt1} and \citep[]{NWeinberg}.

 Using energy conservation law, we set up a first order differential
 equation to describe the evolution of the radius of a cluster
 and solved it in the subsection \ref{MetricAnsaltz}. We then give
 numerical as well as formal analytical solutions for this
 equation. In subsection \ref{KeyParameterSec}, we extract the key
 parameters of this model which will be used in the application of Press-Scheter
 theory. In subsection \ref{theoreticalFormulaSec} we derived
 theoretical formulaes to calculate the number-density v.s.
 temperature of galaxy clusters using Press-Scheter theory. In
 subsection \ref{NRSec}, we provide numerical results for the
 number density of galaxy clusters and its evolutions and
 study the effects of dark energy equation of coefficients on this
 two quantities. We end the paper with the main conclusions and
 some discussions.

 \section{Spherical Collapse Model in QCDM Cosmologies}\label{SCMinQCDM_section}

 \subsection{The Basic Equation and Its Solution}\label{MetricAnsaltz}

 \begin{figure}[t]
 \includegraphics[]{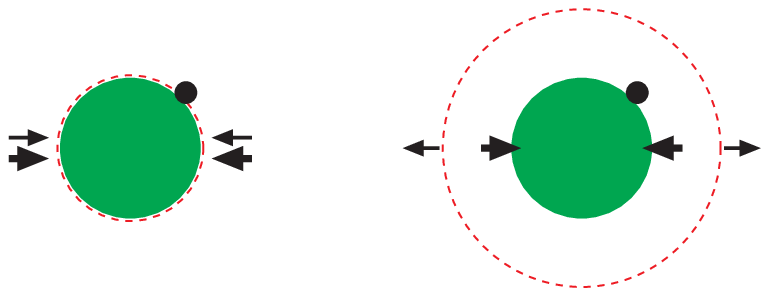}
 \caption{Consider a test particle on the edge of a cluster.
 Left panel, if dark energy cluster synchronously with
 ordinary matters, then energy conservation law will give us a
 standard Friedman equation eq(\ref{standardFriedman}). Right panel,
 if on the scale of galaxy clusters, dark energy does not cluster at
 all, then energy conservation law will also give us a
 Friedmann-like equation, with the potential energy of the test
 particle changed as in eq(\ref{ZGEconserve}).
 }
 \label{EConserve}
 \end{figure}

 There is a simple way \citep[]{SWeinberg} to derive a Friedmann-like equation
 to describe the evolution of the radius of the clusters. Consider
 a test particle which moves synchronously with the the edge of the
 clusters at the gravitation of clusters, see FIG.\ref{EConserve}.
 Denote the scale factor of the background cosmology as $a$,
 while the physical radius of the clusters as $a_p$. At very early times, we can set $a_p\approx a$.
 If the physical radios is taken to be large enough so that it is equal to the hubble length, then
 $a_p=a$ throughout the total evolution process.
 If dark energy
 clusters synchronously with the ordinary matters, just as we assumed
 in \citep[]{SphereI} then according to energy conservation we have
 \beq{}
 \frac{1}{2}m\dot{a}_p^2-\frac{Gm}{a_p}\frac{4\pi a_p^3(\rho_{mc}+\rho_{Qc})}{3}=\textrm{const.}
 \label{standardFriedman}
 \eeq
 which leads to the usual Friedman equation directly. If dark
 energy does not cluster on the scale of galaxy clusters but
 only do so on Hubble scales, just as
 \citep[]{WangSteinhardt1} and \citep[]{NWeinberg} assumed, then
 \beq{}
 &&\frac{1}{2}m\dot{a}_p^2-\frac{Gm}{a_p}\frac{4\pi a_p^3\rho_{mc}}{3}
 \nonumber\\
 &&\hspace{10mm}-\frac{4\pi
 Gm\rho_{Qb}}{3}[\frac{3}{2}a^2-\frac{1}{2}a_p^2]
 =\textrm{const.}
 \label{ZGEconserve}
 \eeq
 which
 can also be written as a Friedman-like form,
 $\dot{a}_p^2=\frac{8\pi
 G}{3}\left[\rho_{mc}a_p^2+\rho_{Qb}(\frac{3}{2}a^2-\frac{1}{2}a_p^2)-k\right]$
 and will become
 eq(\ref{standardFriedman}) in the limit that the radius of the
 clusters becomes the hubble radius.

 It is worth noting that, besides the physical
 radius of the clusters, the symbol $a_p$ appearing in
 eq(\ref{standardFriedman}) can also be understood as
 the scale factor of the clusters-inside space-time.
 But that in eq(\ref{ZGEconserve}) it can only be understood as the
 radius of the clusters, it cannot be understood as the scale
 factor of the cluster-inside space-time. This is because, when
 dark energy is not assumed to move synchronously with ordinary
 matters on the galaxy clusters scale, the metric function
 $U(t,r)$ and $V(t,r)$ in the ansaltz \citep[]{SWeinberg}
 $ds^2=-dt^2+U(t,r)dr^2+V(t,r)(d\theta^2+\textrm{sin}^2\theta
 d\phi^2)$ cannot be factorized $U(t,r)=a^2(t)f(r)$ and
 $V(t,r)=a^2(t)r^2$ uniformly inside the clusters \citep[]{SphereI}.

\begin{figure}[t]
 \centerline{\includegraphics[scale=0.6]{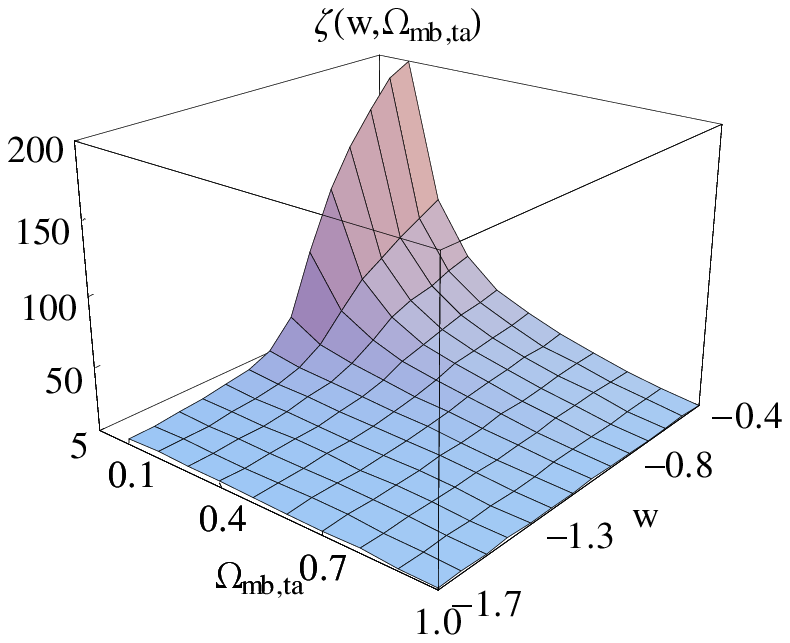}
 \includegraphics[scale=0.6]{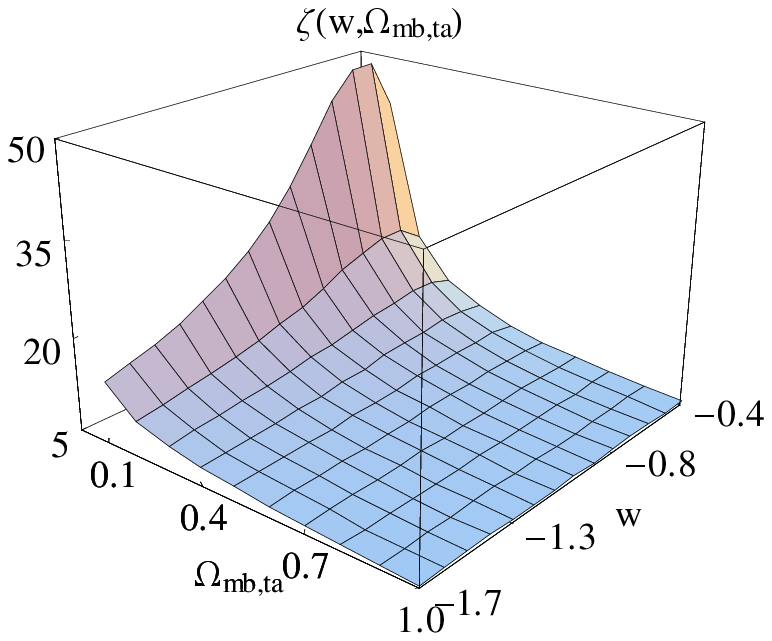}}
 \caption{$\zeta$'s dependence on $w$ and $\Omega_{mb,ta}$.
 Left: dark energy is assumed only to move synchronously with
 ordinary matters on Hubble
 scales, on the galaxy cluster scales, not to cluster at all.
 Right: dark energy is assumed to move synchronously on both
 Hubble scales and galaxy cluster scales.
 }
 \label{zetaFig}
 \end{figure}

 Combining eq(\ref{ZGEconserve})
 with the background universe Friedmann equation:
 \beq{}
 \dot{a}^2=\frac{8\pi G}{3}(\rho_{mb}a^2+\rho_{Qb}a^2)
 \eeq
 we get
 \beq{}
 (\frac{\dot{a}_p}{\dot{a}})^2
 =\frac{\rho_{mc}a_p^2+\rho_{Qb}(\frac{3}{2}a^2-\frac{1}{2}a_p^2)-k}{\rho_{mb}a^2+\rho_{Qb}a^2}
 \label{dapOda}
 \eeq
 using notations
 \beq{}
 &&x=\frac{a}{a_{ta}};\ y=\frac{a_p}{a_{p,ta}};\nonumber\\
 &&\zeta=\frac{\rho_{mc,ta}}{\rho_{mb,ta}};\ \nu=\frac{1-\Omega_{mb,ta}}{\Omega_{mb,ta}}
 \eeq
 eq(\ref{dapOda}) becomes:
 \beq{}
 (\frac{dy}{dx})^2=\frac{\zeta y^{-1}+\nu x^{-3(1+w)}(\frac{3}{2}\zeta^{-\frac{2}{3}}x^2-\frac{1}{2}y^2)-ka_{p,ta}^{-2}}{x^{-1}+\nu x^{-1-3w}}
 \nonumber\\
 \label{yofxDetermination}
 \eeq
 where, as a result of $[\frac{dy}{dx}]_{x=1}=0$,
 \beq{}
 ka_{p,ta}^{-2}=\zeta+\nu(\frac{3}{2}\zeta^{-\frac{2}{3}}-\frac{1}{2})
 \eeq
 Eq(\ref{yofxDetermination}) is a first order non-linear ordinary
 differential equation which contains a characteristic parameter $\zeta$ but
 satisfies a two-boundary condition:
 \beq{}
 y_{x=0}=0;\ y_{x=1}=1
 \eeq
 This kind of equation can be easily solved by the
 numerical method described in \citep[]{PressNR}, chapter 17. We
 display our numerical results in FIG.\ref{zetaFig}. Which is also
 depicted in the figure is the results of
 $\zeta(w,\Omega_{mb,ta})$ when dark energy is assumed to move
 synchronously with ordinary matters both on Hubble scale and
 galaxy cluster scales. The actual fact should lie between this
 two extreme cases. But the result of
 \citep[]{WangSteinhardt1} does not lie between this two extreme
 cases, so it cannot be thought
 as an approximation of the actual facts.

 Eq(\ref{yofxDetermination}) can also be solved analytically. Note
 \beq{}
 \frac{y}{x}=\frac{\frac{a_p}{a_{p,ta}}}{\frac{a}{a_{ta}}}=\frac{a_p}{a}\frac{a_{ta}}{a_{p,ta}}=\frac{a_p}{a}\zeta^{\frac{1}{3}}
 \eeq
 so
 \beq{}
 \left[\frac{y}{x}\right]_{x\rightarrow0}=\zeta^{\frac{1}{3}}\cdot1_-
 \eeq
 Let
 \beq{}
 y=\zeta^{\frac{1}{3}}x\left[1-\alpha (a_{ta}x)+\beta
 (a_{ta}x)^2+...\right]
 \label{yofxAnsaltz}
 \eeq
 substituting eq(\ref{yofxAnsaltz}) into
 (\ref{yofxDetermination}) and equating the two sides of the resulting
 equation order by order in
 $x$, we can get
 \beq{}
 \alpha=\frac{1}{5a_{ta}}\left[\zeta^{\frac{1}{3}}+\nu(\frac{3}{2}\zeta^{-\frac{4}{3}}-\frac{1}{2}\zeta^{-\frac{2}{3}})\right]
 \label{alphaAnalytic}
 \eeq
 and a similar expression for $\beta$, ... etc.
 In this paper we only need to know
 $\alpha$. Some people
 may argue that if progressional solution of
 eq(\ref{yofxDetermination}) is not of the form
 as we wrote in eq(\ref{yofxAnsaltz}) or does not exist at all, then our
 ansaltz eq(\ref{yofxAnsaltz}) will be invalid. We would like to
 point out that if such things occur, then when eq(\ref{yofxAnsaltz})
 is substituted into eq(\ref{yofxDetermination}) we can not get a
 self-consistent equation with the two sides equated order by
 order in $x$.

 \subsection{Extracting The Key Parameters of Press-Scheter Theory from Spherical Collapse
 Model}\label{KeyParameterSec}

 In the ideal model, if there is an over-dense region in a flat
 background universe, then at very early times, this region will
 expand as the background universe expands; but because this
 region's over-dense, its expanding rate will decrease and stop
 doing so at some middle times; then it starts to shrink because
 of
 self-gravitating, the final fate of this over-dense region
 is a singular point. But in practice, when this region shrinks to
 some degree, the pressures originate from the random moving of
 particles inside the over-dense region will balance the
 self-gravitation and the system will enter the virialization
 period. In theoretical studies, it is usually assumed that the
 virialization point is coincident with the collapse point of
 the ideal model on the time axis.

 According to Press-Shceter theory,
 if an over-dense region is to
 be virializated at some time $a_c$, its density-contrast should be
 no less than $\delta_c(w,\Omega_{m0},a_c)$.
 \beq{}
 \delta_c=\left[(\frac{\rho_{mc}(a)}{\rho_{mb}(a)}-1)\frac{1}{D_1(a)}\right]_{a\rightarrow0}D_1(a_c),
 \label{deltacDefinition}
 \eeq
 where $\rho_{mc}$ and $\rho_{mb}$ are the matter densities of
 cluster and background respectively, while $D_1(a)$ is the growth
 function of linear perturbation theory \citep[]{Dodelson},
 \beq{}
 D_1(a)=\frac{5\Omega_{m0}H^2_0}{2}H(a)\int_0^ada^{\prime}[a^{\prime}H(a^{\prime})]^{-3}.
 \eeq
 To reduce the numerical computation burdens, we will use the
 fitting formulaes provided by \citep[]{M99}. We check that when
 $-1.7<w<-0.4$ and $0.1<\Omega_{m0}<0.7$, the formulaes provided by this work is accurate to
 $2\%$.
 To relate the
 mass of a galaxy cluster with its characteristic X-ray temperature,
 the ratio of cluster/background matter densities at
 the virialization point is another very important parameter,
 \beq{}
 \Delta_c(w,\Omega_{m0},a_c)=\frac{\rho_{mc,c}}{\rho_{mb,c}}.
 \label{DeltacDefinition1}
 \eeq

 It can be shown that $D_1(a)_{a\rightarrow0}\rightarrow a$.
 Using eq(\ref{yofxAnsaltz}), we have
 \beq{}
 \left[\frac{a_p}{a}\right]_{a\rightarrow0}=(1-\alpha\cdot a),
 \label{roaagoesto0}
 \eeq
 Substituting
 eq(\ref{roaagoesto0}) into eq(\ref{deltacDefinition}) and using
 the fact that
 $[\frac{\rho_{mc}}{\rho_{mb}}]_{a\rightarrow0}=[\frac{a}{a_p}]^3$,
 we can get
 \beq{}
 &&\hspace{-3mm}\delta_c(w,\Omega_{m0},a_c)=\nonumber\\
 &&\hspace{7mm}\frac{3}{5a_{ta}}
 \left[\zeta^{\frac{1}{3}}+\nu(\frac{3}{2}\zeta^{-\frac{4}{3}}-\frac{1}{2}\zeta^{-\frac{2}{3}})\right]D_1(a_c)
 \nonumber\\
 \eeq

 About $\Delta_c^\prime$s calculation,
 using definition eq(\ref{DeltacDefinition1}) we have
 \beq{}
 \Delta_c(w,\Omega_{m0},a_c)
 =\frac{\rho_{mc,ta}a_{p,ta}^3a_{p,c}^{-3}}{\rho_{mb,ta}a_{ta}^3a_c^{-3}}
 =\zeta\frac{a_{p,ta}^3}{a_{p,c}^{3}}\frac{a_c^3}{a_{ta}^3}
 \label{DeltacDefinition2}
 \eeq
 To calculate the second factor of the above equations' right-most
 part, we can use energy conserving condition and virial theorem.
 If Quintessence clusters synchronously with ordinary matters, by
 assuming that at the collapse point, the system virializes fully,
 we can get the following relations:
 \beq{}
 &&\hspace{-3mm}E_{kinetic,c}=-\frac{1}{2}U_{mm,c}+U_{mQ,c}-\frac{1}{2}U_{QQ,c}
 \label{virialI}\\
 &&\hspace{-3mm}\frac{1}{2}U_{mm,c}+2U_{mQ,c}+\frac{1}{2}U_{QQ,c}\nonumber\\
 &&\hspace{12mm}=U_{mm,ta}+U_{mQ,ta}+U_{QQ,ta}
 \label{EConsI}
 \eeq
 just as we do in \citep[]{SphereI}. Now, since Quintessence is
 assumed not to cluster on the scale of galaxy clusters at all, we
 can only write down the following relations:
 \beq{}
 &&\hspace{-3mm}E_{kinetic,c}=-\frac{1}{2}U_{mm,c}+U_{mQ,c}
 \label{virialII}\\
 &&\hspace{-3mm}\frac{1}{2}U_{mm,c}+2U_{mQ,c}=U_{mm,ta}+U_{mQ,ta}
 \label{EConsII}
 \eeq
 in the above four equations,
 $U_{mm}$, $U_{mQ}$ and $U_{QQ}$ denote the matter-matter,
 matter-Quintessence and Quintessence-Quintessence gravitation
 potentials respectively. The subscripts $_{,c}$ and $_{,ta}$ indicate that
 quantities carrying them should take values at the collapse and turn
 around time respectively. Explicitly,
 \beq{}
 &&\hspace{-3mm}U_{mm,c}=-\frac{3}{5}\frac{GM^2}{a_{p,c}}\propto-\frac{3}{5}\rho_{mc,c}^2a_{p,c}^5\nonumber\\
 &&\hspace{-3mm}U_{mm,ta}=-\frac{3}{5}\frac{GM^2}{a_{p,ta}}\propto-\frac{3}{5}\rho_{mc,ta}^2a_{p,ta}^5\nonumber\\
 &&\hspace{-3mm}U_{QQ,c}\propto-\frac{3}{5}\rho_{Qb,c}^2(a_{p,ta}\frac{a_c}{a_{ta}})^5\\
 &&\hspace{-3mm}U_{QQ,ta}\propto-\frac{3}{5}\rho_{Qb,ta}^2a_{p,ta}^5\nonumber\\
 &&\hspace{-3mm}U_{mQ,c}\propto-[\frac{3}{2}(\frac{a_{p,ta}}{a_{p,c}}\frac{a_c}{a_{ta}})^2-\frac{3}{10}]\rho_{mc,c}\rho_{Qb,c}a_{p,c}^5\nonumber\\
 &&\hspace{-3mm}U_{mQ,ta}\propto-[\frac{3}{2}(\frac{a_{p,ini}}{a_{p,ta}}\frac{a_{ta}}{a_{ini}})^2-\frac{3}{10}]\rho_{mc,ta}\rho_{Qb,ta}a_{p,ta}^5
 \nonumber
 \eeq
 By our normalization, $a_{p,ini}\approx a_{ini}$.
 Substituting these expressions into eq(\ref{EConsII})
 and dividing the resulting equation by the relation
 $\rho_{mc,c}^2a_{p,c}^6=\rho_{mc,ta}^2a_{p,ta}^6$ we get:
 \beq{}
 &&\hspace{-3mm}\frac{a_{p,ta}}{a_{p,c}}
 \left[1+4[\frac{5}{2}(\frac{a_{p,ta}}{a_{p,c}}\frac{a_c}{a_{ta}})^2-\frac{1}{2}]\frac{\rho_{Qb,c}}{\rho_{mc,c}}\right]\nonumber\\
 &&\hspace{15mm}=\left[2+(\frac{5}{2}\frac{a_{ta}^2}{a_{p,ta}^2}-\frac{1}{2})\frac{\rho_{Qb,ta}}{\rho_{mc,ta}}\right]
 \label{aptaOapc_def_QCDM1}
 \eeq
 Using energy conservation law and the approximate mass
 conserving condition,
 $\rho_{mc,c}=\rho_{mc,ta}[\frac{a_{p,ta}}{a_{p,c}}]^3$,
 $\rho_{mc,ta}=\zeta\rho_{mb,ta}$ and
 $\rho_{Qb,c}=\rho_{Qb,ta}[\frac{a_{ta}}{a_c}]^{-3(1+w)}$,
 we can change eq(\ref{aptaOapc_def_QCDM1})
 into the following form:
 \beq{}
 &&\hspace{-3mm}\frac{a_{p,ta}}{a_{p,c}}
 \left[1+4\xi[\frac{5}{2}(\frac{a_{p,ta}}{a_{p,c}}\frac{a_c}{a_{ta}})^2-\frac{1}{2}](\frac{a_{p,ta}}{a_{p,c}})^{-3}
 \right]\nonumber\\
 &&\hspace{29mm}=\left[2+2(\frac{5}{2}\zeta^{\frac{2}{3}}-\frac{1}{2})\nu\zeta^{-1}\right]\nonumber\\
 \label{aptaOapc_def_QCDM2}
 \eeq
 where $\xi=\frac{\nu}{\zeta}[\frac{a_{ta}}{a_c}]^{3(1+w)}$.
 Eq(\ref{aptaOapc_def_QCDM2}) can be solved
 analytically,
 \beq{}
 \frac{a_{p,ta}}{a_{p,c}}&&\hspace{-3mm}=\frac{1}{3}
 \left[-\beta+\frac{\beta^2}{\gamma(\beta,\alpha)}+\gamma(\beta,\alpha)\right]
 \nonumber\\
 \alpha&&\hspace{-3mm}=2\frac{\nu}{\zeta}(\frac{a_{ta}}{a_c})^{3(1+w)}
 \nonumber\\
 \beta&&\hspace{-3mm}=10\frac{\nu}{\zeta}(\frac{a_{ta}}{a_c})^{1+3w}-[2+(5\zeta^{\frac{2}{3}}-1)\frac{\nu}{\zeta}]
 \nonumber\\
 \gamma&&\hspace{-3mm}=\left[\frac{-2\beta^3+27\alpha+\sqrt{27(-4\beta^3\alpha+27\alpha^2)}}{2}\right]^{\frac{1}{3}}
 \nonumber\\
 \label{aptaOapc}\eeq
 Using the fact that
 $t_c=2t_{ta}$ and background Friedman equation $(\frac{\dot{a}}{a})^2\propto(\rho_{mb}+\rho_{Qb})$
 we can set up an integration equation
 \beq{}
 \int_0^{a_c}da^{\prime}\sqrt{\frac{a^{\prime}}{1+\nu_0 a^{\prime
 -3w}}}=2\int_0^{a_{ta}}da^{\prime}\sqrt{\frac{a^{\prime}}{1+\nu_0 a^{\prime -3w}}}
 \nonumber\\
 \label{QCDMtcandtta}
 \eeq
 where $\nu_0=\frac{1-\Omega_{m0}}{\Omega_{m0}}$.
 Solve eq(\ref{QCDMtcandtta}) numerically, we can get the relation $a_{ta}$
 v.s. $a_c$. Substituting eq(\ref{aptaOapc}) and $\frac{a_{ta}}{a_c}$ solved from
 eq(\ref{QCDMtcandtta}) into eq(\ref{DeltacDefinition2}), we will
 finally get the quantity $\Delta_c$.

 \begin{figure}[t]
 \includegraphics[scale=0.6]{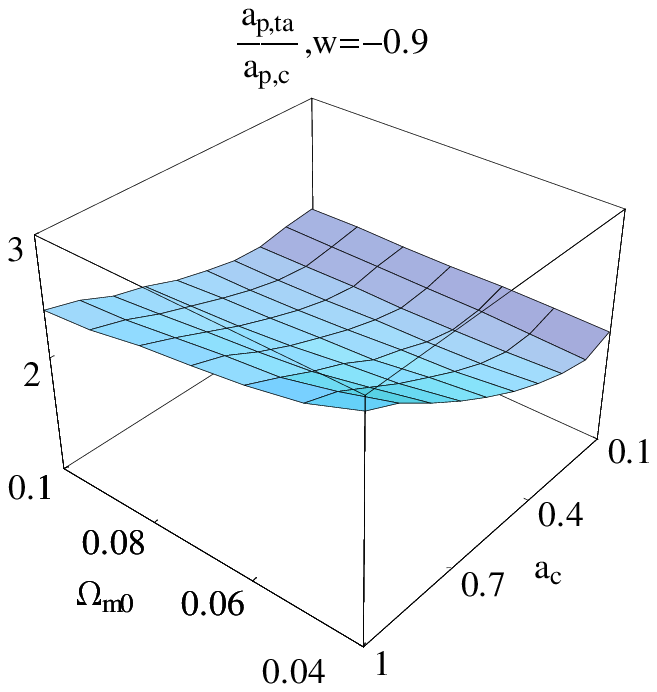}\\
 \centerline{\includegraphics[scale=0.6]{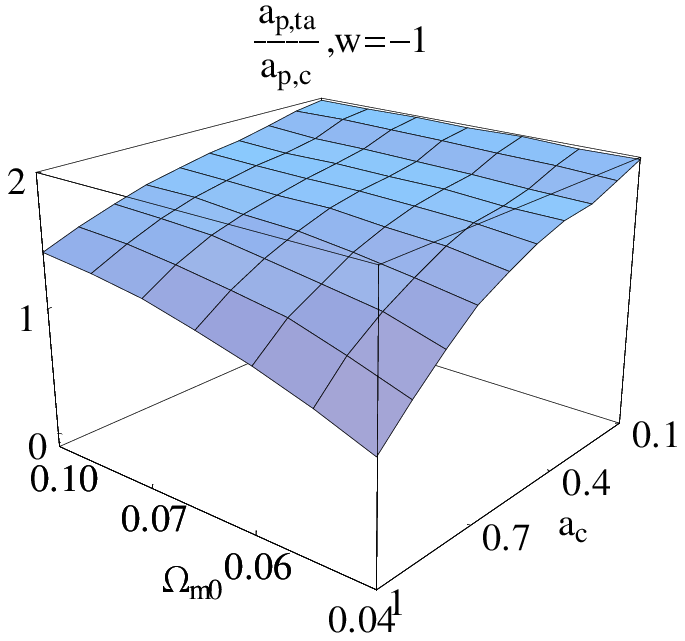}
 \includegraphics[scale=0.6]{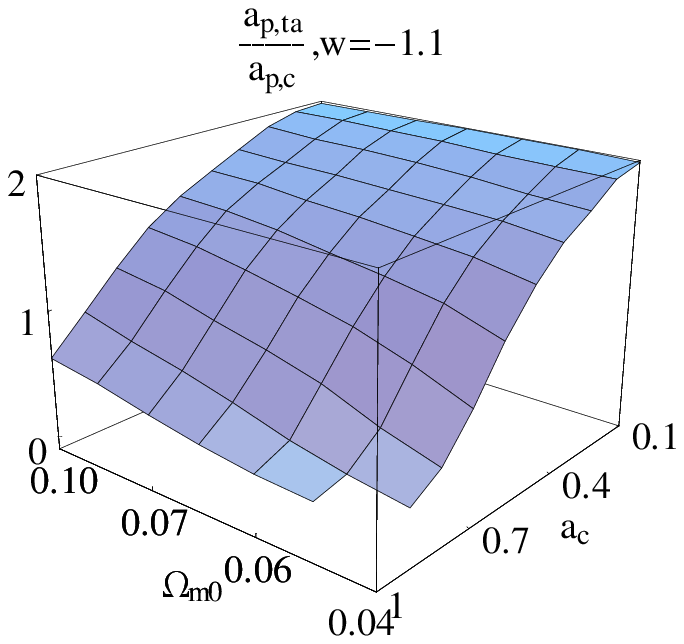}}
 \caption{For too small $\Omega_{m0}$, if a
 cluster is to virialize too later and $w\leq-1$, then at the virialization
 point, its radius will be larger than that at the turn around
 time, i.e. $\frac{a_{p,ta}}{a_{p,c}}<1$.
 }
 \label{aptaOapcFig}
 \end{figure}

 \begin{figure}[t]
 \centerline{\includegraphics[scale=0.6]{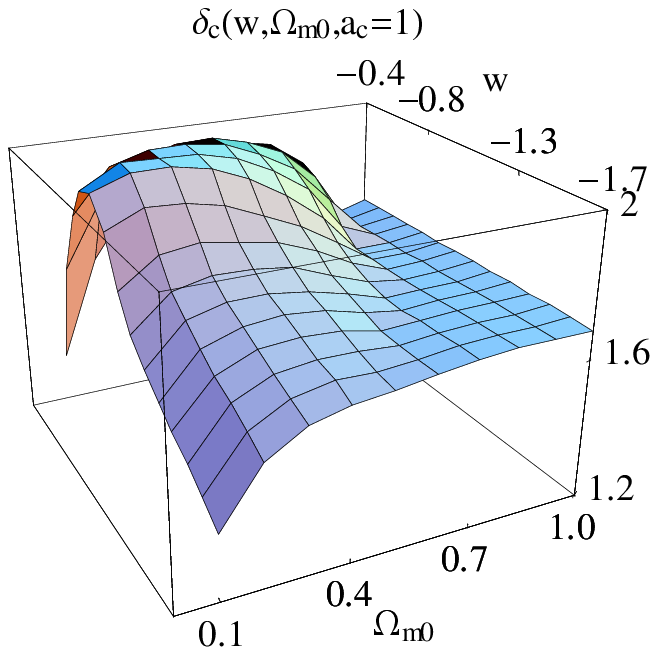}
 \includegraphics[scale=0.6]{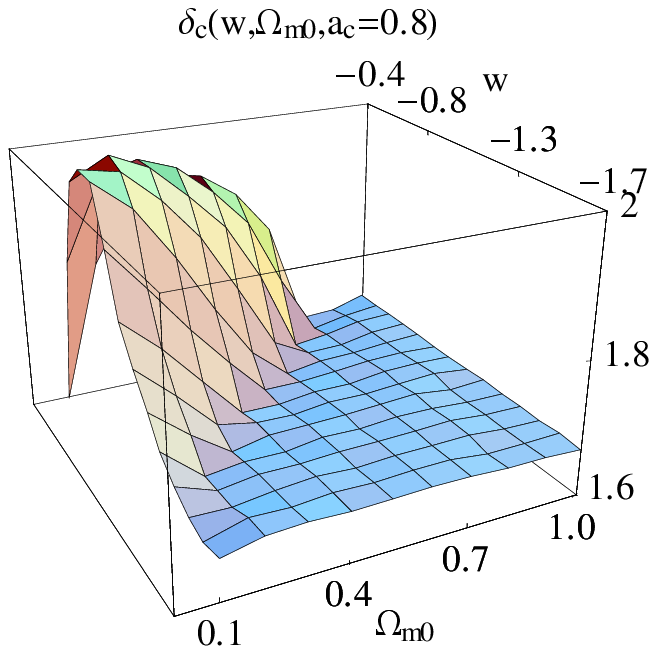}}
 \caption{$\delta_c$'s dependence on $w$, $\Omega_{m0}$ and $a_c$.
 If $a_c\rightarrow0$, $\delta_c\rightarrow1.686$ asymptotically
 whatever $w$ and $\Omega_{m0}$ is.
 }
 \label{dcFig}
 \end{figure}

 \begin{figure}[t]
 \centerline{
 \includegraphics[scale=0.55]{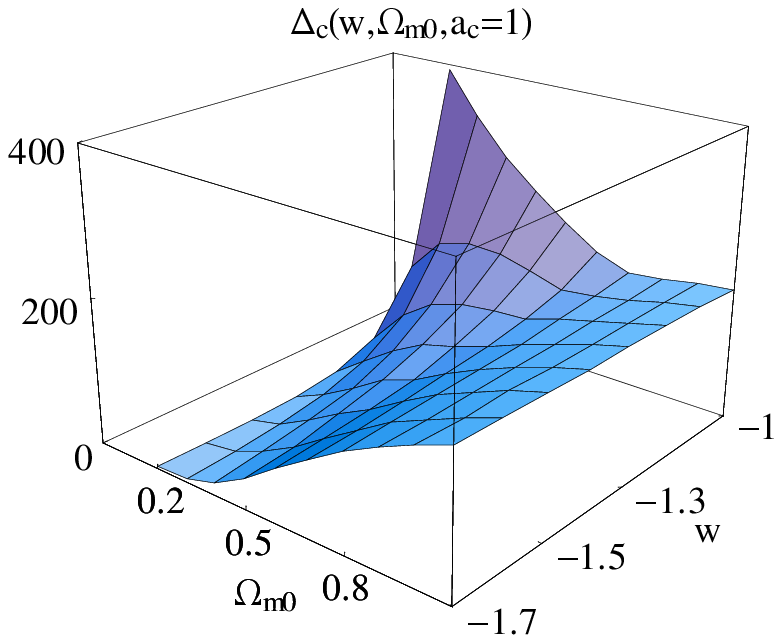}
 \includegraphics[scale=0.6]{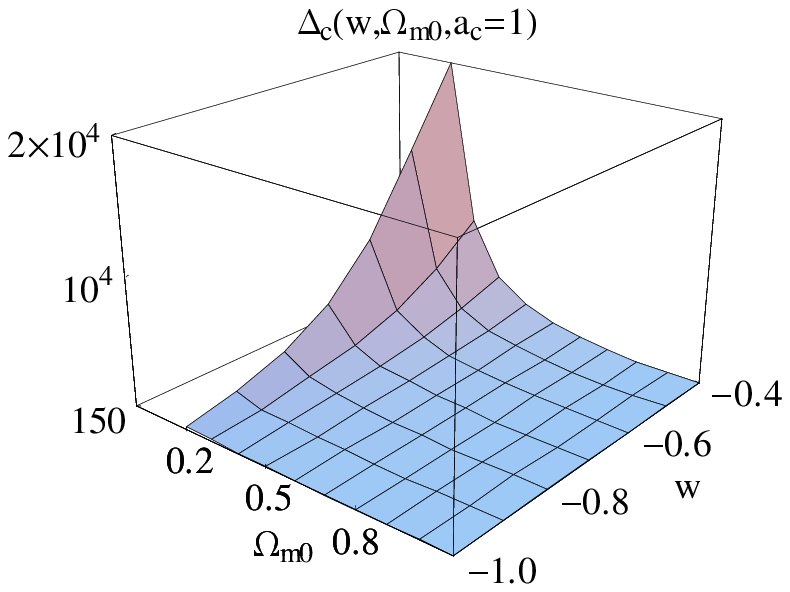}}
 \centerline{
 \includegraphics[scale=0.58]{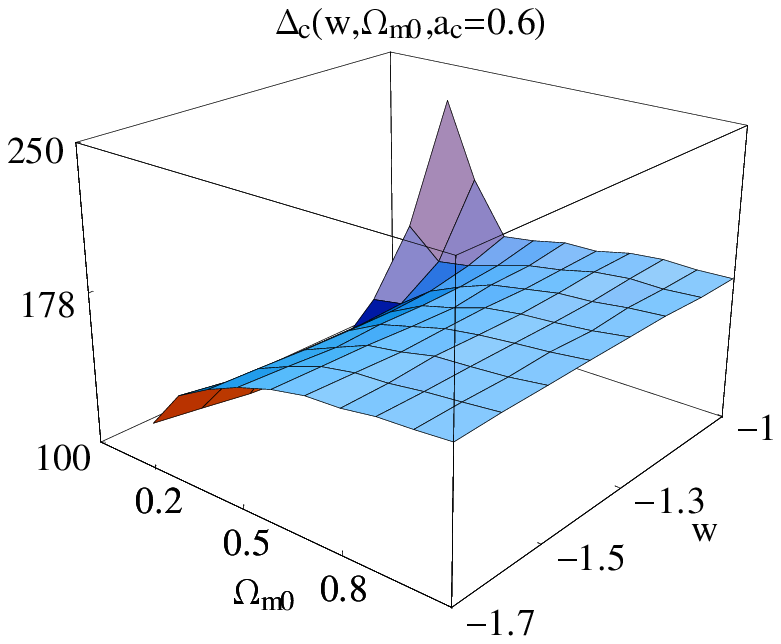}
 \includegraphics[scale=0.6]{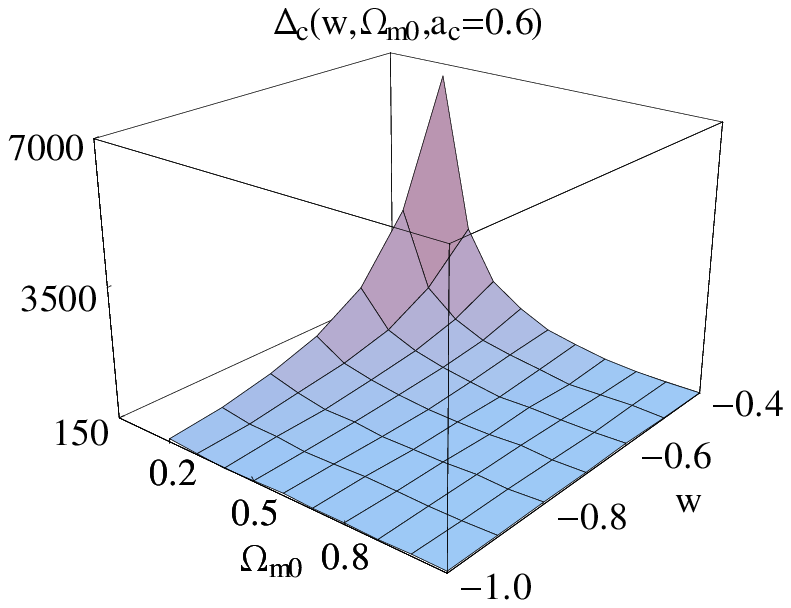}}
 \caption{$\Delta_c$'s dependence on $w$, $\Omega_{m0}$ and $a_c$.
 As $a_c\rightarrow0$, $\Delta_c\rightarrow178$ asymptotically.
 }
 \label{cpDcFig}
 \end{figure}

 We note that, if $w\leq -1$, then for very small $\Omega_{m0}$,
 if a cluster is to virialize too later, then at the virialization
 point, its radius will be larger than that of the turn around
 time, i.e. $\frac{a_{p,ta}}{a_{p,c}}<1$, see FIG.\ref{aptaOapcFig}.
 Physically this means
 that, after the "turn around" point, instead of collapsing, the
 cluster experience a period of expansion to reach virialization
 status. Mathematically
 this only means that the turn around point is a local minimum
 instead of a local maximum of
 the cluster radius and has no problem in principle.
 This may be new structure which has not been
 discovered previously, we call it "phantom hole" and leave
 the detailed discussion of this structure for future works.
 The formation of "phantom hole" will make the kinetic
 energy of the matter-system in it less than $0$, please see
 eq(\ref{vMrelation}), which may be a serious problem.

 We provide numerical results for $\delta_c(w,\Omega_{m0},a_c)$
 and $\Delta_c(w,\Omega_{m0},a_c)$ in FIG.\ref{dcFig}-\ref{cpDcFig}.
 About this two figure, what we would like point out is that, as
 $a_c\rightarrow0$, $\delta_c\rightarrow1.686$ and
 $\Delta_c\rightarrow178$ whatever $w$ and $\Omega_{m0}$ is.
 Physically, this is because, the earlier an over-dense region
 collapse, the more the background universe is like a totally
 matter dominated one. While in a totally matter dominated
 universe, the fact that $\delta_c=1.686$ and $\Delta_c=178$ can
 be proved analytically.

 \section{The Number Density of Galaxy Clusters and Its Evolutions
 }\label{PS_App_Section}

 \subsection{Theoretical Formulaes}\label{theoreticalFormulaSec}

 According to Press-Schechter theory, the comoving number density
 of clusters which have collapsed (i.e., virialized) at certain
 red-shift $z$ and
 have masses in the range $M\sim M+dM$ could be calculated \citep[]{Dodelson}:
 \beq{}
 n(M,z)dM&&\hspace{-3mm}=
 -\sqrt{\frac{2}{\pi}}\frac{\rho_{mb}}{M}
 \frac{\delta_c}{\sigma^2}
 \frac{R}{3M}\frac{d\sigma}{dR}
 \textrm{exp}[-\frac{\delta_c^2}{2\sigma^2}]dM
 \nonumber\\
 \label{diff-M-Function2}
 \eeq
 where: $\rho_{mb}$ is the matter density of background universe;
 the factor $\frac{\rho_{mb}}{M}$ denotes the average number density of
 clusters with mass $M=\frac{4\pi}{3}R^3\rho_{mb}$; $\delta_c$ is given by
 eq(\ref{deltacDefinition}); and
 \beq{}
 &&\hspace{-3mm}\sigma^2(R,z)=\nonumber\\
 &&\sigma_8^2\cdot\frac{\left[\int k^{n_s+2}T^2(k)W^2(k\cdot R)dk\right]D_1^2([1+z]^{-1})}
 {\left[\int k^{n_s+2}T^2(k)W^2(k\cdot 8h^{-1}\textrm{Mpc})dk\right]D_1^2(1)}
 \nonumber\\
 \label{sigmaRz}
 \eeq
 In this paper we use the notations of \citep[]{Dodelson} where $n_s$
 is the primordial power spectrum index, $T(k)$ is the BBKS transfer
 function, $W(k\cdot R)$ is the top-hat window function.

 To relate the mass of a cluster
 with its characteristic X-ray temperature,
 consider a virialized spherical over-dense region
 in the background universe containing dark energies.
 If dark energies cluster synchronously with ordinary
 matters, just as we assumed in \citep[]{SphereI}, then
 we have
 \beq{}
 E_{\textrm{kinetic,vir}}=\left[-\frac{1}{2}U_{mm}+U_{mQ}-\frac{1}{2}U_{QQ}\right]_{,vir}
 \nonumber
 \eeq
 i.e.
 \beq{}
 (\rho_{mc,c}+\rho_{Qc,c})\bar{V}_{vir}^2=\frac{4\pi
 G}{5}\left[a_{p}^2(\rho_{mc}-\rho_{Qc})^2\right]_{,c}
 \nonumber
 \eeq
 where $\bar{V}^2_{vir}$ is the mean
 square velocity of particles in the cluster when the system is
 fully virialized and $a_p$ is the scale factor of the cluster.
 However, if dark energies do not cluster on the scale of
 galaxy clusters at all, then what we can
 get should be
 \beq{}
 &&\hspace{-3mm}\rho_{mc,c}\bar{V}_{vir}^2=\nonumber\\
 &&\hspace{3mm}\frac{4\pi G}{5}a_{p,c}^2\left[\rho_{mc,c}^2
 -[\frac{5}{2}(\frac{a_{p,ta}}{a_{p,c}})^2(\frac{a_c}{a_{ta}})^2-\frac{1}{2}]\rho_{mc,c}\rho_{Qb,c}\right]
 \nonumber
 \eeq
 So
 \beq{}
 &&\hspace{-3mm}\bar{V}_{vir}^2=\frac{3}{5}(GMH)^{\frac{2}{3}}(\frac{\Delta_c}{2})^{\frac{1}{3}}\nonumber\\
 &&\hspace{15mm}\times\left[1
 -[\frac{5}{2}(\frac{\Delta_c}{\zeta})^{\frac{2}{3}}-\frac{1}{2}]\frac{\Omega_{Qb,c}}{\Omega_{mb,c}}\frac{1}{\Delta_c}\right]
 \nonumber\\
 \label{vMrelation}
 \eeq
 where $a_p$ should be understood as the radius of clusters in
 units of Hubble length $H^{-1}$,
 since in this case we cannot define a scale factor globally in
 the clusters and we normalize $a_p$ as $a_p\approx a$ when $a\rightarrow0$. Using relation:
 \beq{}
 k_BT=\frac{\mu
 m_p}{\beta}\frac{\bar{V}^2_{vir}}{3}
 \label{vTrelation}
 \eeq
 where $k_B$ is
 the Boltzmann constant, $m_p$ is the mass of proton, while $\mu m_p$ is
 the average mass of particles, $\beta$ is the ratio of kinetic energy to
 temperature. The composition $\frac{\mu}{\beta}$ has physical meaning of
 energy transformation efficiency from thermal dynamic form to x-ray form.
 Substituting eq(\ref{vMrelation}) into eq(\ref{vTrelation})
 we get the following mass-temperature relation:
 \beq{}
 M=\frac{1}{GH(z)}\left[\frac{5\beta k_BT}{\mu
 m_p}\frac{1}{f(z)}\right]^{\frac{3}{2}}\nonumber
 \eeq
 or
 \beq{}
 R&&\hspace{-3mm}=\left[\frac{2GM}{H^2}\right]^{\frac{1}{3}}
 =\frac{1}{H(z)}\left[\frac{5\cdot2^{\frac{2}{3}}\beta k_BT}{\mu m_p}\frac{1}{f(z)}\right]^{\frac{1}{2}}
 \nonumber\\
 \label{RTrelation}
 \eeq
 with
 \beq{}
 H(z)&&\hspace{-3mm}=H_0[\Omega_{m0}a_c^{-3}+(1-\Omega_{m0})a_c^{-3(1+w)}]^{\frac{1}{2}}
 \nonumber\\
 f(z)&&\hspace{-3mm}=(\frac{\Delta_c}{2})^{\frac{1}{3}}\left[1
 -[\frac{5}{2}(\frac{\Delta_c}{\zeta})^{\frac{2}{3}}-\frac{1}{2}]\frac{\Omega_{Qb,c}}{\Omega_{mb,c}}\frac{1}{\Delta_c}\right]
 \nonumber\\
 \eeq
 and $\Delta_c$ given by eq(\ref{DeltacDefinition2}) and $z=a_c^{-1}-1$.

 Just as \citep[]{WangSteinhardt1} pointed out, since the
 mass-temperature relation is red-shift dependent, simply
 substituting eq(\ref{RTrelation}) into eq(\ref{diff-M-Function2})
 cannot give us correct number density of clusters in a given
 temperature range today. Instead, we should first find out
 the virialization rate and multiply it by the mass-temperature relation
 then integrate over red-shift
 \beq{}
 &&\hspace{-3mm}n(T,z)dT=\nonumber\\
 &&-\frac{1}{\sqrt{2\pi}}\int_z^{\infty}\frac{\rho_{tot}}{MT}
 \frac{d\textrm{ln}\sigma}{d\textrm{ln}R}
 \frac{d\textrm{ln}\sigma}{dz}
 \frac{\delta_c}{\sigma}(\frac{\delta_c^2}{\sigma^2}-1)
 \textrm{exp}[-\frac{\delta_c^2}{2\sigma^2}]dzdT
 \nonumber\\
 \label{diff-T-Function}
 \eeq

 From eqs(\ref{diff-T-Function}), (\ref{RTrelation})
 and (\ref{sigmaRz}) we can see that
 besides $\frac{\mu}{\beta}$,
 $n(T,z)$ will also depend on the cosmological parameters
 $w$, $\Omega_{m0}$, $h$, $n_s$ and the normalization $\sigma_8$
 of the cosmic density fluctuations. In principle, if we can
 measure the number density v.s. temperature relation precisely
 enough, by numerical fittings, we can determine all these parameters
 simultaneously from observations. However, in practice, because
 of parameter degeneracy and measure errors, we can only
 determine some of them or their combinations partly.

 Now let us return to the strange phenomenon displayed in
 FIG.\ref{zetaFig} and \ref{aptaOapcFig}, i.e. $\zeta<1$
 and $\frac{a_{p,ta}}{a_{p,c}}<1$ respectively. We have explained
 that these two things take place when $w<-1$ and
 $\Omega_{mb,ta}$ or $\Omega_{m0}$ takes too small values. It can
 be checked that when $\frac{a_{p,ta}}{a_{p,c}}<1$, we must have
 $\zeta<1$ and the
 kinetic energy of the virialized matter system will be less than
 zero, see eq(\ref{vMrelation}). We think such a "cluster" can
 not emit X-rays.

 \subsection{Numerical Results, Effects of
 $w$ on the Number Density of Galaxy Clusters and
 Its Evolutions}\label{NRSec}

\begin{figure}[t]
 \centerline{
 \includegraphics[scale=0.6]{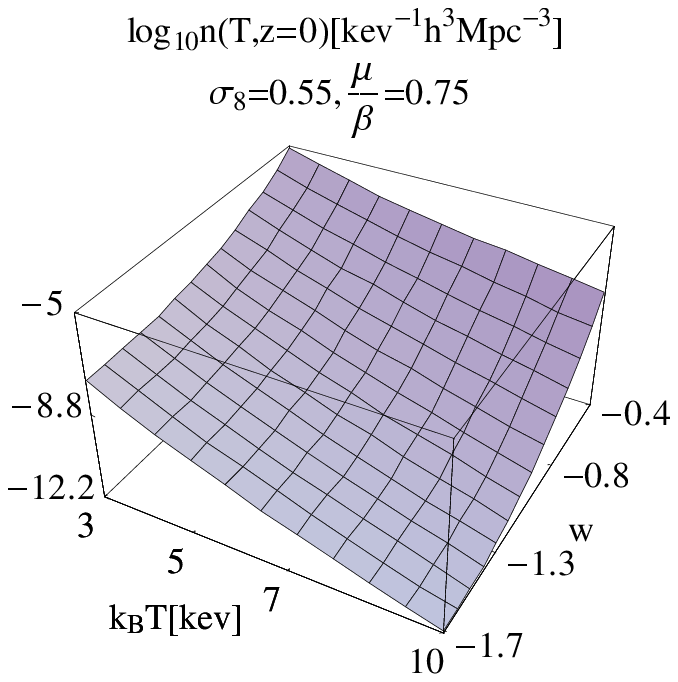}
 \includegraphics[scale=0.6]{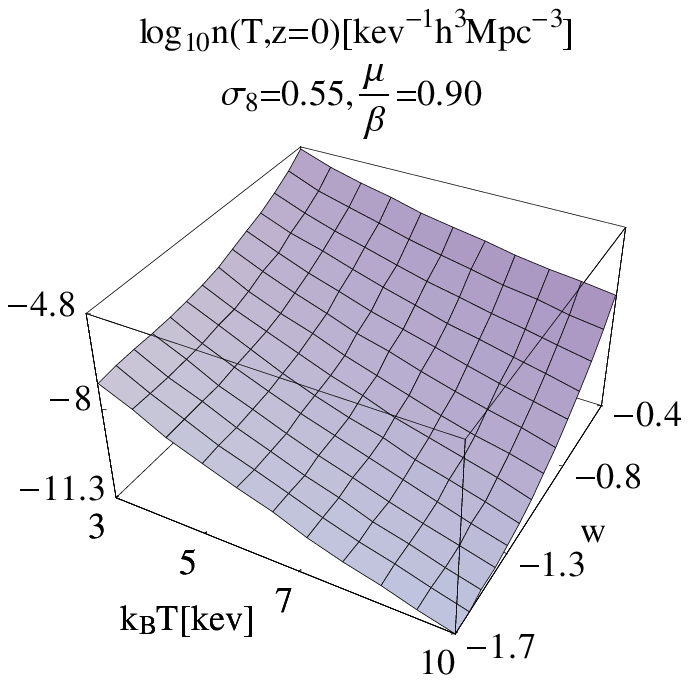}}
 \centerline{
 \includegraphics[scale=0.6]{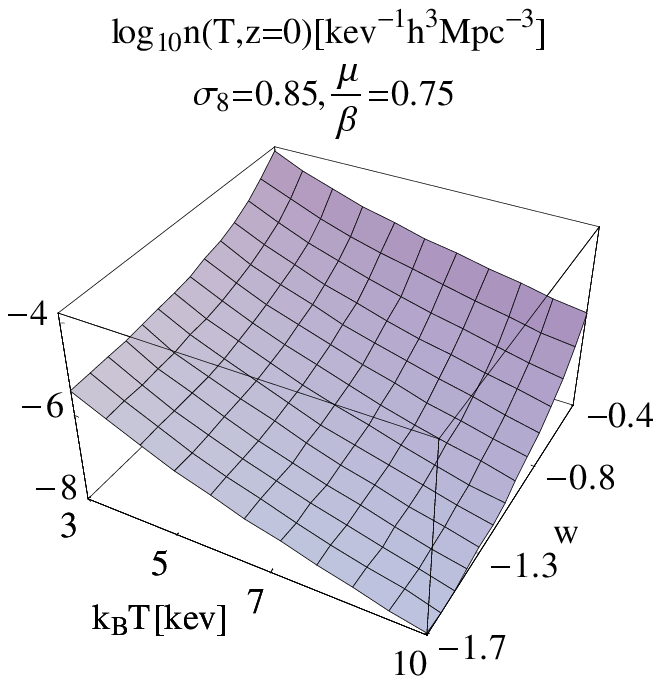}
 \includegraphics[scale=0.6]{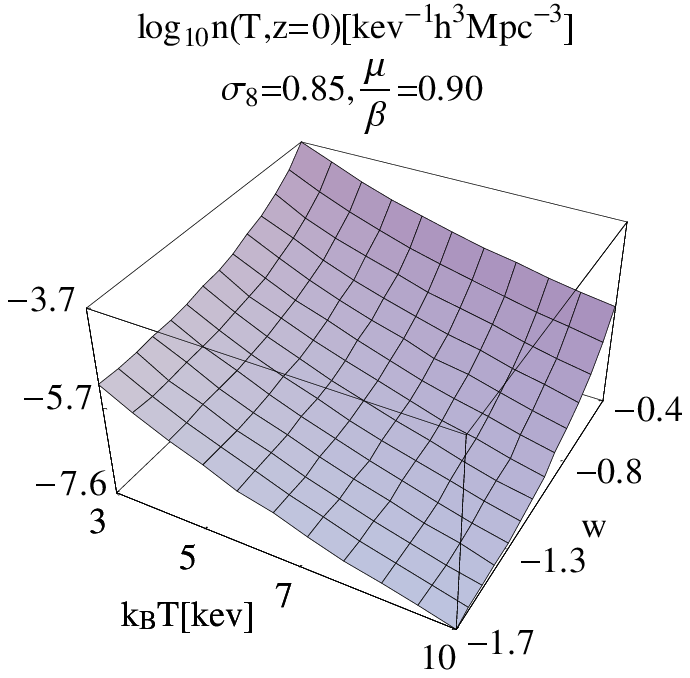}}
 \caption{Effects of $w$ on the
 number density v.s. temperature function  of galaxy clusters when
 $z=0$. The larger is $\sigma_8$ or $\frac{\mu}{\beta}$, the
 larger the function value will be.
 All four figures have
 $\Omega_{m0}=0.27$, $h=0.71$ and $n_s=1.0$.
 }
 \label{ew_nTcore}
 \end{figure}

 We display the effects of $w$ on the number-density v.s.
 temperature of galaxy clusters in FIG.\ref{ew_nTcore} for four
 compositions of $\sigma_8$ and $\frac{\mu}{\beta}$. For any given
 parameter set $\{\sigma_8,\frac{\mu}{\beta}\}$, if $-1<w<0$, then
 $w$ affects the number-density v.s. temperature
 of galaxy clusters exponentially.
 If $w<-1$, the effects are weak.

\begin{figure}[t]
 \centerline{
 \includegraphics[scale=0.6]{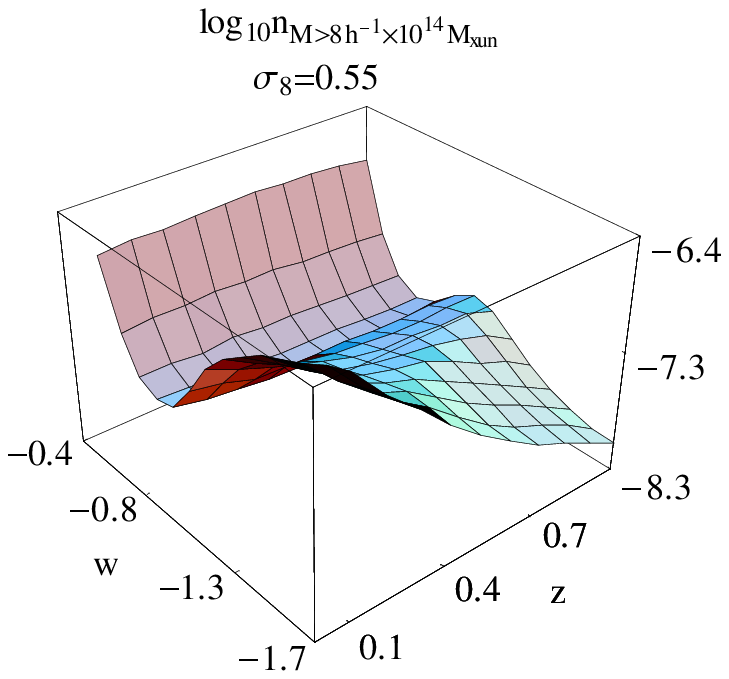}
 \includegraphics[scale=0.6]{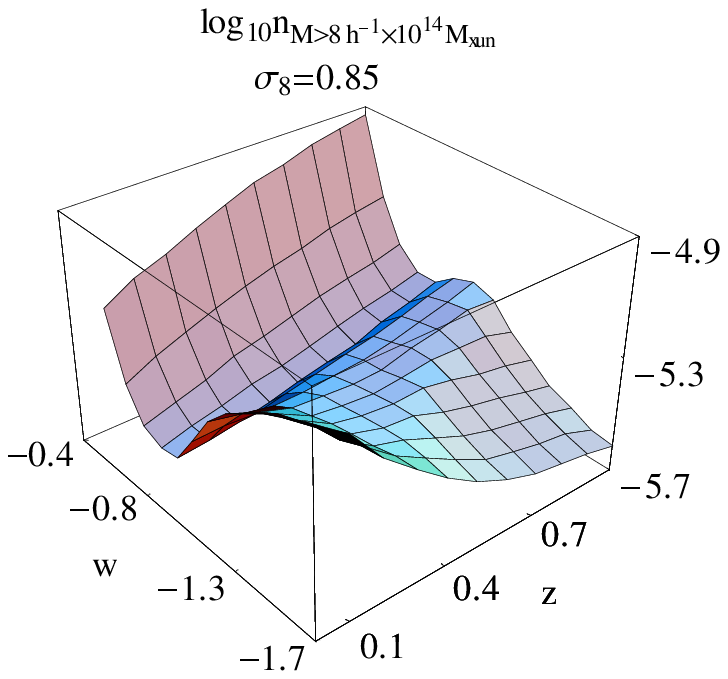}
 }
 \caption{Effects of $w$ on the evolution of number density of
 galaxy clusters whose mass is greater than
 $8h^{-1}\times10^{14}M_{sun}$. When $w$ is greater than $-1$,
 the number density almost does not vary with time for
 $\sigma_8=0.55$, but for $\sigma_8=0.85$ case, the number density
 increases as we look back to the past. When $w$ is less than
 $-1$, the number density decreases at low red-shift, but
 increases at high red-shift.
 Both figure have $\Omega_{m0}=0.27$, $h=0.71$ and $n_s=1.0$.
 }
 \label{ew_nMfunc}
 \end{figure}

\begin{figure}[t]
 \centerline{
 \includegraphics[scale=0.6]{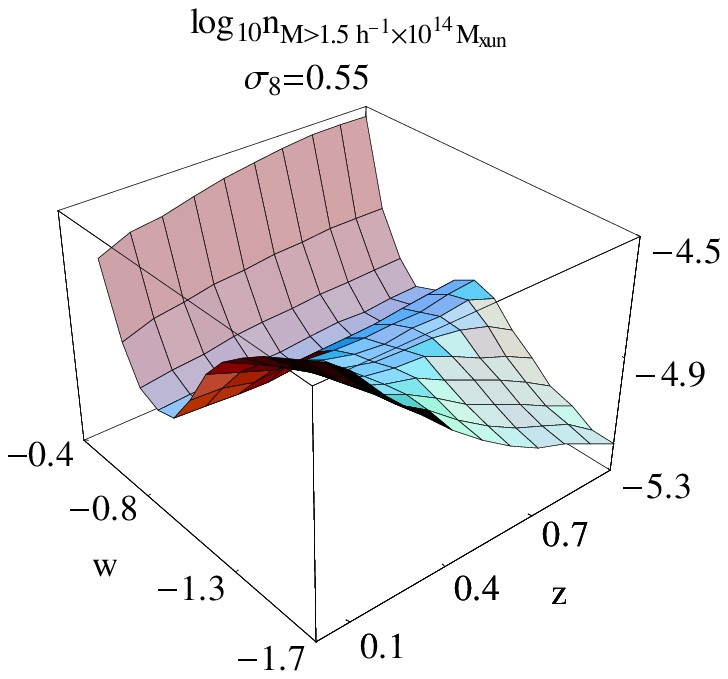}
 \includegraphics[scale=0.6]{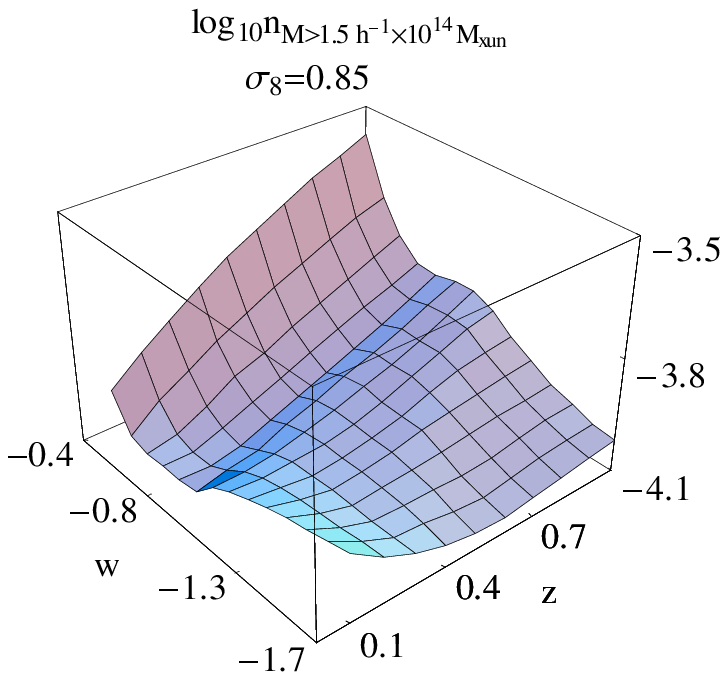}
 }
 \caption{The same as FIG.\ref{ew_nMfunc}, but we consider the
 number density of galaxy clusters whose mass is greater than
 $1.5h^{-1}\times10^{14}M_{sun}$.
 }
 \label{ew_nMfunc2}
 \end{figure}

 In FIG.\ref{ew_nMfunc} we display the effects of $w$ on the
 number-density v.s. red-shift of galaxy clusters whose mass
 is greater than $8h^{-1}\times10^{14}M_{sun}$ for two values of
 $\sigma_8$. From the figure we can easily see that $w$ affects
 the number-density v.s. red-shift relation of galaxy clusters
 remarkably. When $-1<w<0$,
 the number density almost does not vary with time for
 $\sigma_8=0.55$, but for $\sigma_8=0.85$ case, the number density
 increases as we look back to the past. When $w$ is less than
 $-1$, the number density decreases at low red-shift, but
 increases at high red-shift, the turn around red-shift depends on
 $w$. The less is $w$, at the higher red-shift the trend reverses.
 In FIG.\ref{ew_nMfunc2}, we depict the same effects for the
 number density of galaxy clusters whose mass is greater than
 $1.5h^{-1}\times10^{14}M_{sun}$. From the figure, we can see for
 the smaller mass galaxy
 clusters, the effects of $w$ on the number-density v.s. red-shift
 is more remarkable.

 Comparing this fact with the observational results reported by
 \citep[]{BFC97} and \citep[]{BahcallBode03}, we can almost immediately
 say that $w$ can not be greater than $-1$! Of
 course, a reliable conclusion should be obtained by best
 fitting the current observational results with theoretical
 predictions. But if we want to fit the results of \citep[]{Edge,HA91,Henry97}
 with our theoretical formulae eq(\ref{diff-T-Function}), we at
 least have four parameters $\frac{\mu}{\beta}$, $w$, $\sigma_8$ and
 $\Omega_{m0}$ to determine. If we want to fit the results of
 \citep[]{BahcallBode03}, we have to treat the problem of fitting
 data with errors in both coordinates. This two kinds of
 operations both cost time formidably. We leave them for
 future works.

 Comparing the results in this paper with that of \citep[]{SphereI},
 we can see that the effects of $w$ on the evolution of number density of galaxy
 clusters is even more strong under the assumption that dark
 energy does not cluster on the scale of galaxy clusters than the
 case where dark energy is assumed to cluster synchronously with
 ordinary matters. It's easy to imagine that,
 the actual case should lie between
 this two extreme way, matter clusters and forms potential well, then dark
 energy falls into it and cannot climb up so clusters also.
 If we want to fit observation results
 into theoretical predictions to get reasonable constraints on
 $w$, $\Omega_{m0}$, $\sigma_8$ and other cosmological parameters,
 theoretical formulaes in the both extreme cases are needed, and
 some kind of interpolation should be used.

 Whatever the actual case is, according to the results
 found in this paper and
 its previous counterpart \citep[]{SphereI}, we know that the
 effect of $w$ on the number density of galaxy clusters
 is so strong that we
 think it should be possible to use this effect to measure $w$.

 \section{Conclusions}\label{ConSection}

 We study the top-hat spherical collapse model of galaxy clusters
 formation in the flat QCDM or Phantom-CDM cosmologies under the
 assumption that Quintessence or Phantom does not cluster
 on this scale. We find that under this
 assumption, the key parameters of the model exhibit rather
 non-trivial and remarkable dependence on the equation of state
 coefficients $w$ of Quintessence or Phantoms. We then applied the
 results in Press-Scheter theory and calculated the number density
 v.s. temperature function and the evolution
 of the number density of massive galaxy clusters and find that these two
 Quantities are both affected by $w$ exponentially.

 For the number density v.s. temperature function
 of galaxy clusters, we find that it is an increasing function of
 $w$ and the dependence on $w$ is more strong in the range
 $-1<w<0$ than it is in the range $-\infty<w<-1$.
 While for the evolution of the number density
 of massive galaxy clusters, we find that when $w$ is less than
 $-1$, the number density decreases at low red-shift, but
 increases at high red-shift, the turn around red-shift depends on
 $w$. The smaller is the galaxy clusters' mass, the stronger is
 the this effect.
 According to the observational result, which says that
 the number density of massive galaxy clusters decreases
 as we look back to the past, we can
 qualitatively conclude that $w$
 should not take values greater than $-1$ too much. It should take values
 less than $-1$.

 The actual dark energy
 cluster property should lie between the two extreme cases discussed in this paper
 and that in \citep{SphereI}. But whatever the fact is,
 our results here and that in \citep{SphereI} indicate that,
 $w$ affects the number density of galaxy clusters exponentially.
 So we think measuring the number density of galaxy clusters and its
 evolutions may be an effective way to determine $w$.

 As discussions, we would like to state that, if the problem we pointed out
 in \citep{SphereI} is the fact, then we may have to accept that, we ignored
 a very important assumption in the usual $\Lambda$CDM cosmology. That is:
 the cosmic component denoted by $\Lambda$ and leading to the accleration of the
 universe moves synchronously with ordinary matters. The reason is very clear,
 if $\Lambda$ does not move synchronously with ordinary matters on
 Hubble scales, then in our comoving reference frame build on the
 ordinary matters(such as supernovaes), we should have a $\Lambda$ current
 flowing outside the hubble horizon. Once that occurs, we cannot define
 an all universe uniformly defined scale factor, so we have no Friedmann
 equation at all. In that case, our current explanation of
 the acceleration indicated by the observation
 of supernovaes will be very problematic.

 If $\Lambda$ component moves synchronously with ordinary matters, then we will
 have no reason to say that it is the vacuum energy of quantum field.
 On the contrary, some kinds of couplings between $\Lambda$ (or dark energy) and
 the ordinary matters is a must be derivation.

 \acknowledgements

 Part of the numerical computations are performed on the parallel
 computers of the Inter-discipline Center of Theoretical Studies
 of ITP, CAS, Beijing, China.

 \begin{center}
 {\bf Appendix. Comparing With Previous Works}
 \end{center}
 \begin{center}
 {\bf A. Comparing The Basic Equations}
 \end{center}

 Our results in this paper and its sibling
 one \citep{SphereI} are so different from the previous
 works that without a concrete comparison and explicit pointing
 out the problem in those works, few peoples will believe our
 conclusions.

 To compare our basic equations with the basic equations used by
 \citep{WangSteinhardt1}, we can differentiate
 eq(\ref{ZGEconserve}) and use energy conservation law to get,
 \beq{}
 \frac{\ddot{a}_p}{a_p}=-4\pi G\left[
 (w+\frac{1}{3})\rho_{Qb}\frac{3}{2}\frac{\dot{a}}{\dot{a_p}}\frac{a}{a_p}
 +\frac{1}{6}\frac{(\rho_{Qb}a_p^2)^\cdot}{a_p\dot{a_p}}+\frac{1}{3}\rho_{mc}
 \right]\nonumber\\
 \label{WSbasicEq1}
 \eeq
 In the right hand side of this equation,
 if we let $a==a_p$, i.e., assume that the dark energy moves synchronously
 with ordinary matters on the galaxy cluster scales, then it reduce to the
 eq(A2) of \citep{WangSteinhardt1}. Note, in the right hand side of
 eq(\ref{WSbasicEq1}), $a$ only appears in terms where dark energy
 is involved. But
 \citep{WangSteinhardt1} does not use this assumption consistently,
 because when it combine its eq(A2) with the equation describing
 the dark energy's evolution eq(A6), it uses the assumption that only on Hubble
 scales, dark energy moves synchronously with ordinary matters.

 Although we have pointed out in \citep{SphereI}, we still would like to
 point out that, as long as variable separation technique is used
 in solving Einstein equation, then whichever (Raychaudhuri or Freidmann) equation we choose
 to describe the evolution of the over-dense region, we will in fact
 have assumed that dark energy moves synchronously with ordinary
 matters in our over-dense region. Let us explain this point in
 more details. For an over-dense collapsing region, the most
 general metric describe its inside space-time is
 \beq{}
 ds^2=-dt^2+U(t,r)dr^2+V(t,r)(d\theta^2+\textrm{sin}^2\theta
 d\phi^2),\nonumber\\
 \eeq
 Using Einstein equation $G_{\mu\nu}=-8\pi GT_{\mu\nu}$, we can
 prove that, only when no energy current flowing outside the
 over-dense region, is the energy momentum tensor $T_{\mu\nu}$
 diagonal, and can the metric function $U(t,r)$ and $V(t,r)$ be
 factorized as
 \beq{}
 U(t,r)=a_p^2(t)f(r),\ V(t,r)=a_p^2(t)r^2.
 \eeq
 And only when the $U(t,r)$ and $V(t,r)$ function is
 factorized, can we have
 \beq{}
 &&\hspace{-3mm}\frac{\ddot{a}_p}{a_p}=-\frac{4\pi
 G}{3}[\rho_{mc}+\rho_{Qc}+3p_{Qc}],
 \label{time_time_Einstein}
 \\
 &&\hspace{-3mm}\textrm{in (WS98)'s notation}
 \nonumber
 \\
 &&\hspace{-3mm}\frac{\ddot{R}}{R}=-4\pi G(p_Q+\frac{\rho_Q+\rho_{cluster}}{3})
 \nonumber
 \eeq
 \beq{}
 \frac{\ddot{a}_p}{a_p}+2\frac{\dot{a}_p^2}{a_p^2}+2\frac{k}{a_p^2}=4\pi
 G[\rho_{mc}+\rho_{Qc}-p_{Qc}].\label{space_space_Einstein}
 \eeq
 As a must be of eqs(\ref{time_time_Einstein}) and (\ref{space_space_Einstein})
 \beq{}
 \frac{\dot{a}_p^2}{a_p^2}+\frac{k}{a_p^2}=\frac{8\pi G}{3}(\rho_{mc}+\rho_{Qc})
 \eeq
 So Freidmann equation and Raychaudhuri equation must hold at
 the same time, or must not hold simultaneously. The statement of
 \citep{WangSteinhardt1} and \citep{NWeinberg} that when dark energy
 does not cluster on the scale of galaxy clusters, Raychaudhuri
 equation can be used to describe the evolution of the over-dense
 region but Freidmann equation does not hold is an incorrect
 statement.

 Let us say more explicitly, when dark energy does not cluster on
 the galaxy clusters, hence a dark energy exists which describe the
 flowing of dark energy outside the over-dense region, the
 basic equation which should be used to describe the evolution of
 the over-dense region is not the eq(A2) of
 \citep{WangSteinhardt1}, it should be our eq(\ref{ZGEconserve}).
 Our eq(\ref{ZGEconserve}) is not obtained by variable separation
 in solving Einstein equation. We obtained it by energy
 conservation, so we include the effects of the dark energy
 current on the evolution of the over-dense region.

 \begin{center}
 {\bf B. Comparing The Numerical Results}
 \end{center}

 Let us emphasize again that the problem in the existing works
 \citep{WangSteinhardt1} and \citep[]{NWeinberg} is, when writing
 down the basic equations describing the evolution of the
 over-dense region they assumed that dark energy moves
 synchronously with ordinary matters, please see
 eq(A2) of \citep{WangSteinhardt1}, but when writing down
 equations which will be used to describe the
 density of dark energies in the over-dense regions, please see eq(A6) of
 \citep{WangSteinhardt1}, a different assumption is made. That is,
 dark energy only moves synchronously with ordinary matters on
 Hubble scales.

 Just as we pointed out in \citep[]{SphereI} and in the
 conclusion section of this paper. In realities, dark energy
 should have some degree of cluster behaviors on the galaxy
 clusters. We can imagine, ordinary matters cluster and form
 potential wells, when dark energy falls in and some
 degree of dark energy's clustering will occur either. So the
 actual case of dark energy's clustering phenomenon should lie
 between the following two extreme cases. The two extreme cases
 are, dark energy moves synchronously with ordinary matters on
 both Hubble scales and galaxy cluster scales or dark energy only
 moves synchronously with ordinary matters on Hubble scale but
 could not fall in the potential wells formed by the over-dense
 matter region at all.

\begin{figure}[t]
 \includegraphics[scale=0.45]{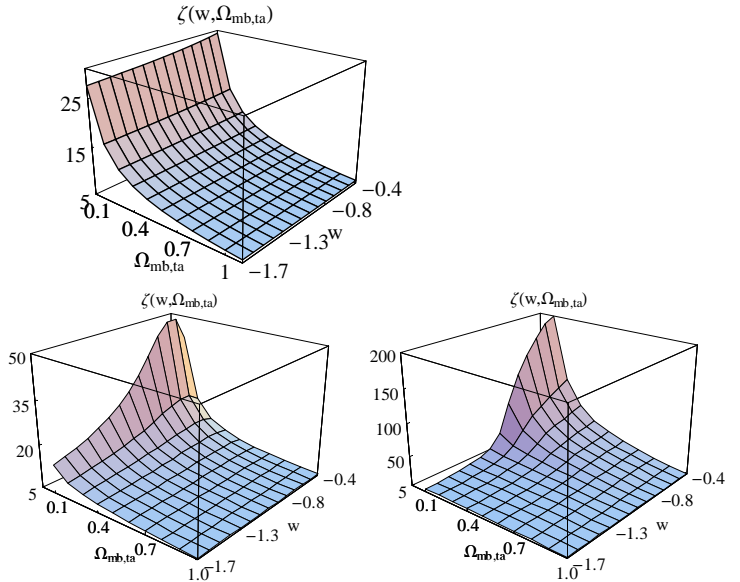}
 \\
 \centerline{
 \includegraphics[scale=0.6]{Fig2b.eps}
 \includegraphics[scale=0.6]{Fig2a.eps}
 }
 \caption{
 $\zeta^\prime$s dependence on $w$ and
 $\Omega_{mb,ta}$. The upper, solved from
 eq(A2) and (A6) of \citep[]{WangSteinhardt1}, when the basic
 equation (A2) were written to describe the evolution of the over-dense
 regions, the assumption that dark energy moves synchronously with
 ordinary matters was made, but when the dark energy density was
 involved in the right hand side of the equation, a different
 assumption is made. That is, only on Hubble scales, dark energy moves
 synchronously with ordinary matters. Down left, dark energy is
 assumed to move synchronously with ordinary matters both on
 Hubble scales and on galaxy clusters scale. Down right, dark
 energy is only assumed moves synchronously with ordinary matters
 on Hubble scales, it is not assumed to move synchronously with
 ordinary matters on galaxy clusters scale.
 }
 \label{zetaCompare}
 \end{figure}

\begin{figure}[t]
 \includegraphics[scale=0.45]{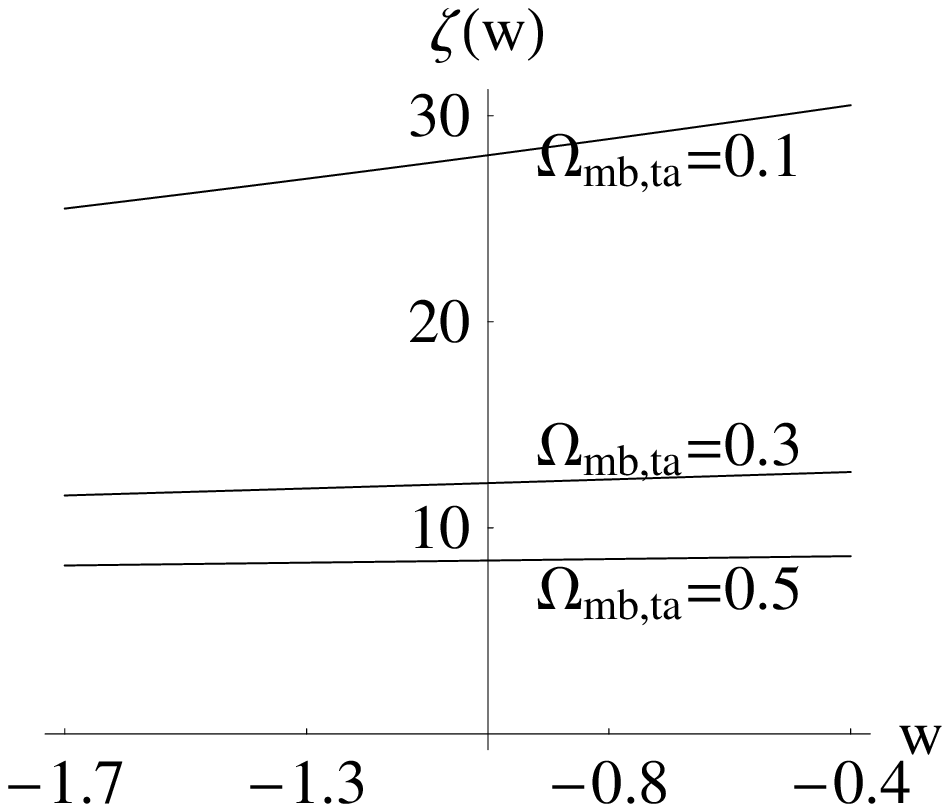}
 \centerline{
 \includegraphics[scale=0.45]{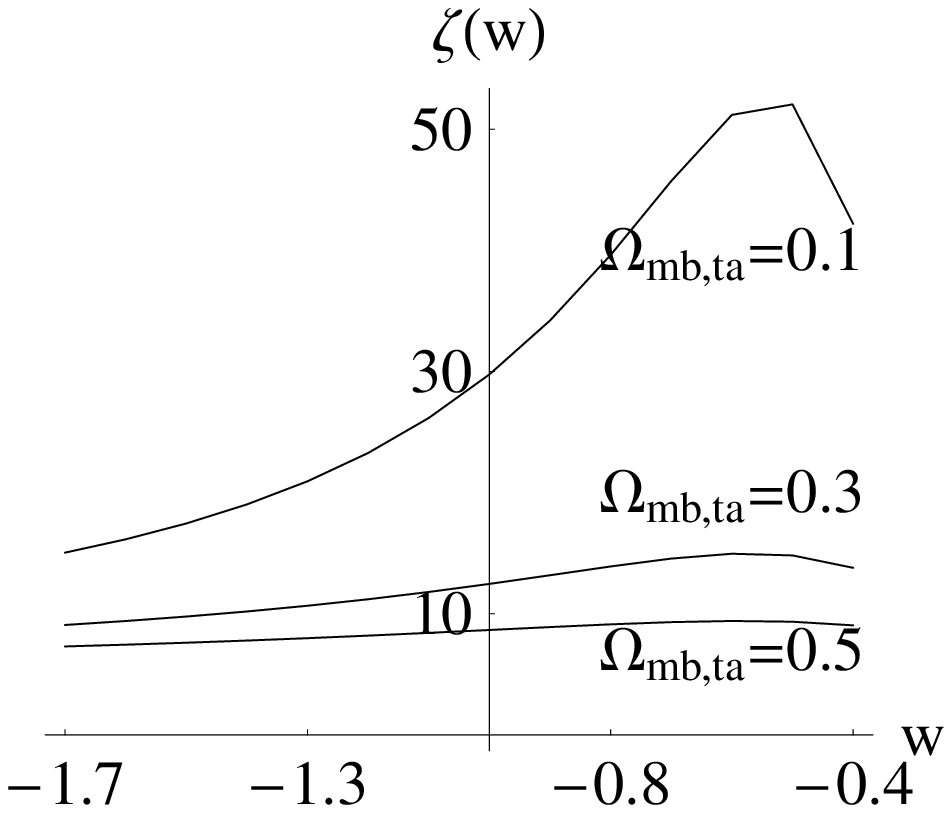}
 \includegraphics[scale=0.45]{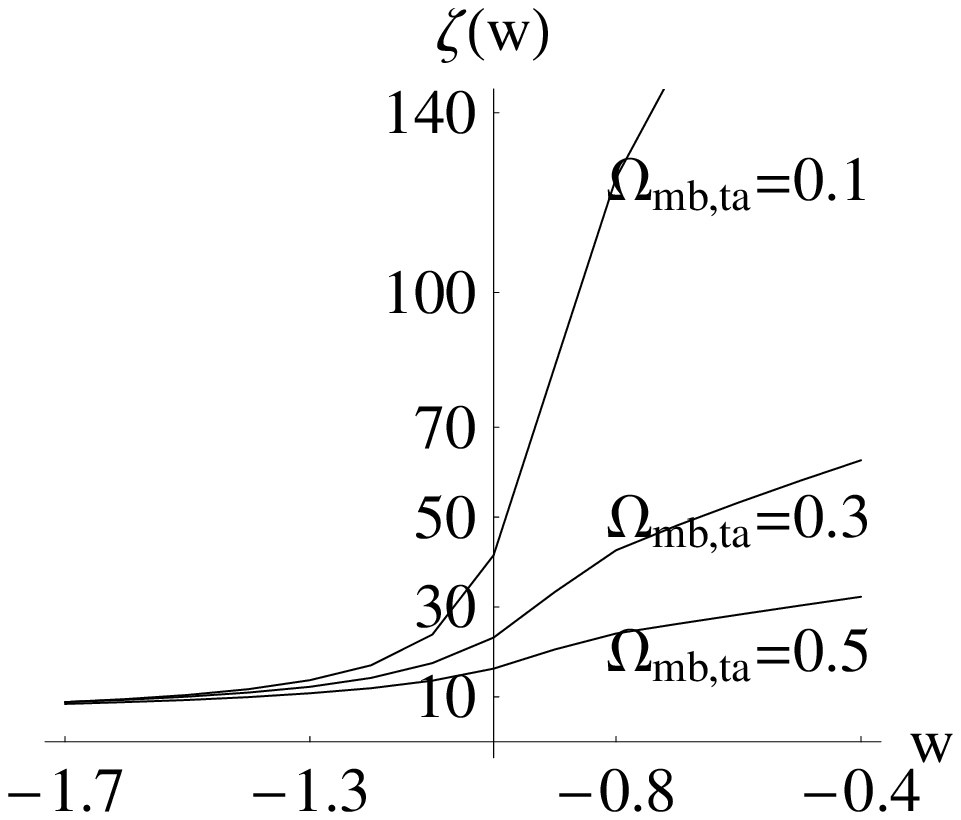}
 }
 \caption{
 The same as FIG.\ref{zetaCompare}, but $\Omega_{mb,ta}$ is set to
 three special value $0.1,\ 0.3,\ 0.5$.
 }
 \label{zetaCompare2d}
 \end{figure}

 We study the first extreme case in \citep[]{SphereI} and the second
 extreme case in this paper. A natural question is, can the results
 of \citep[]{WangSteinhardt1} lie between our two extreme cases? If
 this is the case, then although inconsistent, the treatment of
 \citep[]{WangSteinhardt1} can be thought as some kinds of
 approximation of realities. We will see in the following that
 this is not the case. We compared the results of $\zeta^\prime$s
 dependence on $w$ and $\Omega_{mb,ta}$ in FIG.\ref{zetaCompare}
 and \ref{zetaCompare2d}, FIG.\ref{zetaCompare} is 3-dimensional,
 FIG.\ref{zetaCompare2d} is 2-dimensional.

 \begin{center}
 {\bf C. Some Comments On References}
 \end{center}

 When this paper and its sibling one \citep[]{SphereI} are put on the e-preprint
 arXive and submitted to Astrophysical Journals, we are told that
 just recently, many authors have studied this
 problem in different depth. Such as \citep[]{BM05}, \citep[]{BW03}, \citep[]{Koi05},
 \citep[]{MB04}, \citep[]{NM04} and \citep[]{NM05}. We believe there
 must be more we do not know at this moment. We think we
 should first
 express our thanks to the referee and editors of APJ and the
 authors of these papers for their informing us of these works, we
 then also would like to point out that, almost all these works contain the
 problem we pointed out in this paper, i.e., when writing down the
 basic equations to describe the evolution of the over-dense
 regions, the assumption that dark energy moves synchronously with
 ordinary matters is made, but in the basic equations when the
 dark energy's density is involved, the results from another
 different assumption is used.


\begin{thebibliography}{}

\bibitem[Bahcall, Fan \& Cen(1997)]{BFC97} Bahcall, N. A., Fan, X.H. \& Cen, R. Y., 1997, \aj, 485, L53

\bibitem[Bahcall \& Bode(2003)]{BahcallBode03} Bahcall, N. A. \& Bode, P., 2003, \aj, 588, L1-L4

\bibitem[Bardeen et al(1986)]{BBKS} Bardeen J. M., Bond J. R., Kaiser N. \& Szalay A. S., 1986, \aj, 304, 15

\bibitem[Barrow \& Saich(1991)]{Barrow} Barrow, J. D. \& Saich, Paul, 1991, \mnras, 262, 717

\bibitem[van de Bruck \& Mota(2005)]{BM05} van de Bruck, C. and Mota, D.F., Invited talk at IDM 2004: 5th
International Workshop on the Identification of Dark Matter,
Edinburgh, Scotland, United Kingdom, 6-10 Sep 2004,
astro-ph/0501276.

\bibitem[Battye \& Weller(2003)]{BW03} Battye, R.A. \& and Weller, J., 2003, PRD, 68, 083506

\bibitem[Dodelson(2003)]{Dodelson} Scott Dodelson, 2003, Modern Cosmology, Academic Press

\bibitem[Edge et al(1990)]{Edge} Edge A. C., Stewart G.C., Fabian A. C. \& Arnaud K. A., 1990, \mnras, 245, 559

\bibitem[Eke, Cole \& Frenk(1996)]{ECF} Eke, V. R., Cole, S. \& Frenk,C. S., 1996, \mnras, 282, 263

\bibitem[Hengry \& Arnaud(1991)]{HA91} Henry J. P. \& Arnaud K. A., 1991, \aj, 372, 410

\bibitem[Henry(1997)]{Henry97} Henry J. P., 1997, \aj, 489, L1

\bibitem[Koivisto(2005)]{Koi05} Koivisto, T., astro-ph/0504571.

\bibitem[Lokas \& Hoffman(2001)]{LokasHoffman} Lokas,E. L., and Hoffman, Y., astro-ph/0011295, astro-ph/0108283.

\bibitem[Ma et al(1999)]{M99} Ma, C. P., Caldwell, R. R., Bode, P. \& Wang L., 1999, \aj, 521, L1

\bibitem[Mota \& van de Bruck(2004)]{MB04} Mota, D.F. and van de Bruck, C., 2004, Astron. Astrophys.  421, 71

\bibitem[Nunes \& Mota(2004)]{NM04} Nunes, N. and Mota, D.F., Structure Formation in Inhomogeneous Dark Energy Models, astro-ph/0409481

\bibitem[Manera \& Mota(2005)]{NM05} Manera, M., Mota, D.
 Cluster number counts dependence on dark energy inhomogeneities and
 coupling to dark matter, astro-ph/0504519

\bibitem[Press et al(1992)]{PressNR} W. H., Presset al. 1992, Numerical Recipes, Cambridge University Press.

\bibitem[Percival, Miller \& Peacock(2000)]{Peacock} Percival, W. J., Miller, L., Peacock, J.A., 2000, \mnras, 318, 273

\bibitem[Wang \& Stein hardt(1998)]{WangSteinhardt1} Wang, Li-Min, Steinhardt, P. J., 1998, \aj, 508,483

\bibitem[Weinberg(1972)]{SWeinberg} Weinberg, S., 1972, Gravitations and Cosmology, John Wiley press.

\bibitem[Weinber \& Kamionkowski(2002)]{NWeinberg} Weinberg, N. N. and Kamionkowski,M, 2002, \mnras, 000, 1

\bibitem[Zeng \& Gao(2005)]{SphereI} Zeng, D. F. and Gao, Y. H., Spherical Collapse Model and Dark Energy(I), astro-ph/0505163.

\end{thebibliography}
\end{document}